# Simulating Acculturation Dynamics Between Migrants and Locals in Relation to Network Formation

Rocco Paolillo

Bremen International Graduate School of Social Sciences

University of Bremen & Jacobs University Bremen

Germany

Wander Jager

Groningen Center for Social Complexity Studies

University College Groningen, University of Groningen

the Netherlands

Abstract

International migration implies the coexistence of different ethnic and cultural groups in the receiving country. The refugee crisis of 2015 has resulted in critical levels of opinion polarization on the question of whether to welcome migrants, causing clashes in receiving countries. This scenario emphasizes the need to better understand the dynamics of mutual adaptation between locals and migrants, and the conditions that favor successful integration. Agent-based simulations can help achieve this goal. In this work, we introduce our model MigrAgent and our preliminary results. The model synthesizes the dynamics of migration intake and post-migration adaptation. It explores the different acculturation outcomes that can emerge from the mutual adaptation of a migrant population and a local population depending on their degree of tolerance. With parameter sweeping, we detect how different acculturation strategies can coexist in a society and in different degrees among various subgroups. The results show higher polarization effects between a local population and a migrant population for fast intake conditions. When migrant intake is slow, transitory conditions between acculturation outcomes emerge for subgroups, e.g., from assimilation to integration for liberal migrants and from marginalization to separation for conservative migrants. Relative group sizes due to speed of intake cause counterintuitive scenarios, such as the separation of liberal locals. We qualitatively compare the processes of our model with the German portion sample of the survey "Causes and Consequences of Socio-Cultural Integration Processes among New Immigrants in Europe" (SCIP), finding preliminary confirmation of our assumptions and results.

Keyword list: acculturation, migration, tolerance, polarization, agent-based simulation

Introduction

International migration implies individuals or entire populations crossing international boundaries (International Organization for Migration [IOM], 2017). It is a complex phenomenon common in human history that connects people from different cultures and historical backgrounds. Understanding the consequences of international migration for receiving societies has become more important considering the increase in migration flows to Europe and other global destinations since 2015, namely due to the refugee crisis (European Commission [EC], 2015; Organization for Economic Co-operation and Development [OECD], 2015). At the peak of the crisis, as reported by the European Commission (2015), the United Nations High Commissioner for Refugees [UNHCR] confirmed that 1,015,078 migrants had reached Europe via the Mediterranean Sea in 2015. As Eurostat (2017) has reported, EU countries approved 307,650 asylum requests during 2015, and the number increased in 2016 to 672,890 requests accepted. In 2016, Germany approved 433,905 requests, while Sweden approved 66,585 and Italy 35,405. Although the peak of the refugee crisis is likely in the past (UNHCR, 2017), factors such as climate change might increase migration pressure in the future due to floods, droughts and associated conflicts. For example, Kelley, Mohtadi, Cane, Saeger and Kushnir (2015) have linked the conflict in Syria and the associated refugee crisis to unprecedented droughts in the region.

The consequences of such migration for receiving countries are evident. A survey by the Pew Research Center (Wike, Stokes & Simmons, 2016) has demonstrated opinion polarization in European countries due to the refugee crisis. For example, the discussions on Brexit focused on, among other topics, the sovereignty of the UK in the context of refugee acceptance. On the one hand, many Europeans consider refugees as unwilling to integrate and as representing a threat to the stability and economy of their nations (Wike et al., 2016). On the other hand, many support the urgent calls to open borders and accept refugees as a form of humanitarian aid (Wike et al., 2016). This internal chasm regarding the acceptance of migrants is more evident than at earlier stages of European history (OECD, 2015). Additionally, many asylum seekers who have been granted the status of humanitarian migrants are likely to settle in Europe in the long term. As social scientists, we are interested in the consequences of this scenario as regards the cohesion of the receiving societies and the integration of migrants.

Facing these issues requires one to consider the dynamics of acculturation in the context of continuous first-hand contact between members of different cultures (Redfield, Linton, & Herskovits, 1936). A prominent model in the field of acculturation studies is the fourfold model proposed by Berry (1984, 2005), which identifies acculturation as a mutually adaptive process between a local community and migrant minorities. It assumes that acculturation occurs at two levels: (1) a cultural-group level made up of norms and (2) a psychological-individual level concerning the individual experience of adaptation (Berry, 1984, 2005). The model is based on the orientations people adopt to face issues related to maintaining their own cultural identities and participating in the larger society. Four orientations emerge from the intersection of the two distinct dimensions:

- Integration: high maintenance of one's own culture, frequent interactions with the other group
- Assimilation: low maintenance of one's own culture, frequent interactions with the other group
- Separation: high maintenance of one's own culture, infrequent interactions with the other group
- Marginalization: low maintenance of one's own culture, infrequent interactions with the other group

The most recent elaborations of the model have associated societal ideologies with the different orientations (Berry, 2005, 2017). Integration is associated with multicultural societies and people able to navigate the features of different cultures. Assimilation entails melting pot policies and incorporation within the receiving society. Separation means segregation between ethnic groups, while marginalization is associated with exclusion. Regardless of the original focus on the orientations, Berry's categories are universally recognized as a typological model describing the range of possibilities regarding how people situate themselves and participate in different cultural groups (Nguyen & Benet-Martinez, 2013).

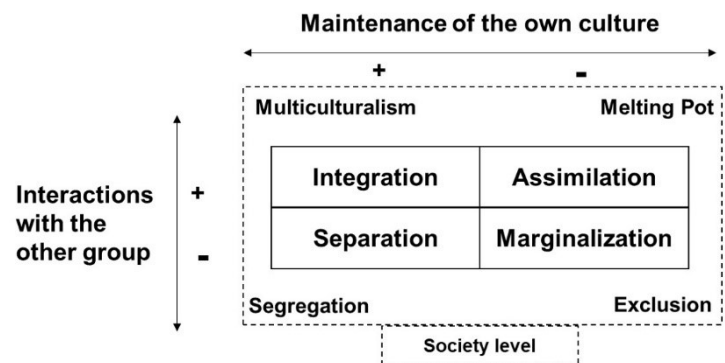

Figure 1. Acculturation orientations and societal profiles. Adapted from Berry (2005)

Many social science studies have explored what processes favor one strategy over another and how desired integration outcomes can be reached (Ward & Kus, 2012). The dynamics of acculturation are influenced by many factors. In this paper, we focus on the role of tolerance for both the receiving community and migrants. Apparently simple in its meaning, the term tolerance, when applied to integration in a diverse context, is of interest in domains ranging from social science research (Brewer & Pierce, 2005) to policy design (Verkuyten, Yogeeswaran, & Adelman, 2018). In this context, tolerance implies acceptance despite recognition and disapproval of diversity (van Doorn, 2014). Nevertheless, as Verkuyten et al. (2018) have highlighted, in most social sciences fields, tolerance is synonymous with openness to cultural others or a generalized "positive attitude" (p.10) towards them. Both meaning highlight the role of tolerance as an antecedent of network formation in a multicultural context due to migration flows. Roccas and Brewer (2002) have demonstrated that low levels of tolerance towards outgroups (e.g., migrants and those of different race) are associated with conservatism values, which motivate people to avoid uncertainty and ambiguity. On the other hand, tolerance can cause group cohesion to fracture. On a personal level, tolerant people might prefer tolerant people of other groups over conservative members of their own group (Verkuyten, 2010).

Conversely, conservative people might reject tolerant members of their same group for holding different values (Verkuyten, 2010). Moreover, conservative people might openly condemn the acceptance of cultural others by members of their own group out of a sense of ethnic loyalty (Padilla & Perez, 2003).

Even if tolerance is accepted as a possible predictor of migrant integration, an important question is how it actually modulates the mutual adaptation between locals and migrants. The contact hypothesis, stating how positive contact between groups reduces prejudice, can fit this interest (Allport, 1954). Although this assumption is not completely new within acculturation studies (Berry, 1984), it has received more emphasis in recent years (Berry, 2017). Even though reciprocity in social interactions is recognized as a fundamental element of network formations and their consolidation (Stark, 2015), the dynamics of the process are complex and conclusions not obvious. Research has clearly demonstrated that positive intergroup interaction reduces prejudice towards outgroups (Dovidio, Eller, & Hewstone, 2011). Nevertheless, Riek, Mania and Gaertner (2006) have shown that negative intergroup experiences, either real or imagined, increase negative attitudes towards outgroups. Due to this complexity, many scholars have highlighted the need to more effectively address the interactions occurring in the ecological contexts of acculturation and the ways in which they condition integration processes (Ward & Geeraert, 2016; Berry, 2017).

We believe that knowledge of how tolerance influences the social dynamics of mutual adaptation between locals and migrants in relation to network formation is crucial for understanding how polarizing opinions on out-groups can influence future integration scenarios. Agent-based modelling can be useful in this regard for several reasons. First, agent-based modelling allows researchers to translate their own hypotheses on social processes into distributed actions of individuals in artificial societies (Macy & Willer, 2002). For this reason, agent-based modelling is the only method able to simulate and explore the emergence of complex social phenomena linking the micro-level of individual behavior with the macro-level of observed social phenomena (Edmonds & Meyer, 2017; Gilbert & Troitzsch, 2005). Second, model building allows us to compare different types of artificial societies, so to embrace the complexity necessary to disentangle conditions and processes underlying social phenomena such as the contact hypothesis. Lastly, what-if scenarios allow us to connect different processes that can contribute to integration outcomes in a sequential order, a feat that variable-based models are not capable of (Squazzoni, Jager & Edmonds, 2013). As in the case of acculturation studies, the clarification of mechanisms requires examining the relationship among migration flows, speed of intake, network formation and changes in tolerance as an effect of reciprocity.

In sum, following Edmonds (2017), we propose using agent-based modelling to explain some plausible social mechanisms underlying the emergence of acculturation within Berry's typological model in a migration context. To that end, we developed the MigrAgent model, which this paper presents. In our model, we focus on the role of tolerance and reciprocity in intergroup network formation. In the first experiments described in this paper, we compare different societies in terms of their level of conservatism and simulate migration flows using varying speed of intakes. Our aim is to determine what type of acculturation emerges for different strata of society from these scenarios. In this preliminary work, we qualitatively compare the assumptions of our model and observed processes with data from the survey *Causes and Consequences of Socio-Cultural Integration Processes among New Immigrants in Europe* (SCIP) collected in Germany.

Model Description

MigrAgent was built using NetLogo 6.04, and it is currently available at the CoMSES Computational Model Library, see section Additional Information. In the model, the world is split between a home country and a host country. Local agents (blue color tag) reside in the host country, and migrant agents (green color tag) reside in the home country. Once the simulation starts, migrant agents move from the home country to the host country according to speed of intake, which represent the random probability of an agent to move to the host country. No agent, whether local or migrant, can leave the host country. The three main dynamics of the simulation occur in the host country in a circular way. The first dynamic is group aggregation based on agents' attachment preferences; together with the speed of intake, this variable influences spatial sorting and the probability of interaction between agents. The second dynamic is the acceptance or rejection of intergroup interactions based on the tolerance of the receiver. The last dynamic is the change in tolerance as a function of reciprocity (see Figure 2). Table 1 summarizes the parameters and the attributes of the agents.

Following Berry's description of acculturation as involving group-level norms (2005), we included a group-level dimension of conservatism to simulate tolerance. The collective conservatism of each ethnic group ranges from -1 and +1. The individual level of conservatism of each agent at the start of the simulation runs (time step 0) is determined according to a normal distribution with mean equal to the collective conservatism of the own ethnic group and standard deviation of 0.45. The individual level of conservatism of each agent then changes as a dynamic variable according to their experience of rejection or acceptance in intergroup interactions, as described below. Values of individual conservatism below 0 denote liberal agents (bright color tag); values of conservatism equal to or higher than 0 denote conservative agents (dark color tag). In using the terms *conservative* and *liberal* agents, we do not refer to any political or philosophical interpretation, but to agents with a low level of tolerance (conservatives) or a high level of tolerance (liberals). Thus, four types of agents interact in our simulation: liberal locals (bright blue), conservative locals (dark blue), liberal migrants (bright green) and conservative migrants (dark green). Ethnicity is a fixed state, and conservatism is a dynamic state.

| Parameter | Type | Range | Function |
|---|---|---|---|
| number_local | integer | [0,1000] | number of local agents |
| conservatism_local | continuous | [-1,1] | mean in normal distribution of local agents' conservatism. standard deviation = 0.45 |
| number_migrant | integer | [0,1000] | number of migrant agents |
| conservatism_migrant | continuous | [-1,1] | mean in normal distribution of migrant agents' conservatism. standard deviation = 0.45 |
| speed_intake | integer | [1,100] | speed of intake: percentage of migrant agents moving simultaneously to host country |
| Agents attributes | Type | Range | Tag and Action |
| ethnicity | fixed | [local, migrant] | blue color: local green color: migrant |
| conservatism | dynamic | [-∞,+∞] | liberal agents [< 0]: bright ethnicity color conservative agents [≥ 0]: dark ethnicity color |
| happy? | boolean | [true, false] | liberal agents: fraction liberal agents in-radius 1.5 ≥ 0.5 conservative agents: fraction conservative same ethnicity in-radius 1.5 ≥ 0.5 |

Table 1. Parameters of MigrAgent model

We designed our model based on Schelling's dynamics of segregation (1971), although we did not literally copy the rules of his model. We took the core assumption that individual homophily preferences can lead to high levels of segregation. Although Schelling's model is usually applied to spatial segregation, we applied it to network aggregation. Our idea was that people's segregation under certain homophily preferences limits individuals' opportunities to interact with members of other groups, regardless of their willingness to engage in such interactions. In our adaptation, liberal agents consider as similar other agents based on their attitude, not their ethnicity. They prefer for their proximal neighborhood to be home to agents with low conservatism, independent of their ethnicity, and for that neighborhood to exclude conservative agents, including those of their own group. Conservative agents instead consider as similar agents sharing both their ethnicity and their high level of conservatism. Thus, conservative agents reject both liberals of their own ethnic group and agents of different ethnicities. They prefer being in neighborhoods hosting only conservative agents of their own ethnicity. In Schelling's original model, agents are *happy* with their neighborhood when it includes a certain fraction of desired agents ranging from 0 to 1. To control parameter sweeping, we kept the desired fraction of agents fixed at 0.5. This means that agents are happy with their neighborhood if those considered as similar are not in a minority condition; moreover, an agent could be happy if alone. Neighborhood space is calculated as an in-radius of 1.5. With Schelling's dynamics, agents do not move if a satisfactory proportion of desired agents live in their neighborhood. We differentiated our model in this regard. In our simulation, agents move within a neighborhood if they are satisfied with its composition; if dissatisfied, they relocate elsewhere. We made this change to increase the likelihood of intergroup interaction and to simulate how people's sense of security (Berry, 2017) influences their exploration of the world. Due to this decision, we do not envision the model achieving a stable equilibrium, and the model does not feature a stop rule. The formula below represents the behavior of agents:

$$f_i^C = \frac{\sum_{i \in y}(C \subset E_i)}{N_{i \in y}} \qquad f_i^L = \frac{\sum_{i \in y}(L)}{N_{i \in y}}$$

where:

$f_i^C$ = fitness of conservative agent $i$

$f_i^L$ = fitness of liberal agent $i$

$i \in y$ = neighborhood $y$ of agent $i$

$C \subset E_I$ = conservative agents of the ethnic group $E$ of agent $i$

$L$ = liberal agents of both ethnic groups

$N_{i \in y}$ = total number of agents in neighborhood $y$ of agent $i$

If $f_i \geq 0.5$, agent $i$ moves within the neighborhood

If $f_i < 0.5$, agent $i$ relocates elsewhere

Following Berry's definition of a psychological-individual level of adaptation in acculturation processes (2005), the initial conservatism of agents varies as a function of reciprocity or rejection in intergroup interactions. The segregation dynamics in the host country that have been described influence the probability of interaction. At each step, agents propose an interaction with others in the neighborhood (in-radius 1.5) regardless of their ethnicity or conservatism. In making this choice, we assumed that any person is potentially connected with others in his or her spatial proximity. Ingroup interactions occur by default; intergroup interactions are accepted by liberal agents and rejected by conservative agents. In our view, this assumption distinguishes between the behavior expected from liberal versus conservative agents in an intercultural context. The individual level of conservatism of each agent increases following their experiences of intergroup rejection and decreases due to intergroup acceptance. For experiences of both acceptance and rejection, the highest value is kept in memory is as follows:

$$c_{i,t} = c_{i,t-1} + max_i^{rej} - max_i^{acc}$$

where:
$c_{i,t}$ = conservatism of agent *i* at time step *t*
$c_{i,t-1}$ = conservatism of agent *i* at time step *t-1*
$max_i^{rej}$ = maximal experience of rejection of agent *i*
$max_i^{acc}$ = maximal experience of acceptance of agent *i*

Note that $c_{i,t}$ at time step 0 follows the distribution of collective conservatism of ethnic group of agent *i*

Once agents are no longer in the same neighborhood due to their movements, the local interaction breaks, although the maximal experience of rejection and maximal experience of acceptance are kept in the memory of the agent.
Based on the dynamics thus described, the acculturation categories in Berry's model are defined as the types of direct links that an agent receives; these links simulate participation in each ethnic group (see Figure 3):

- Integration: The agent receives interactions from the ingroup and out-group
- Assimilation: The agent receives interactions from the out-group, but not from the ingroup
- Separation: The agent receives interactions from the ingroup, but not from the out-group
- Marginalization: The agent receives interactions from neither of the groups

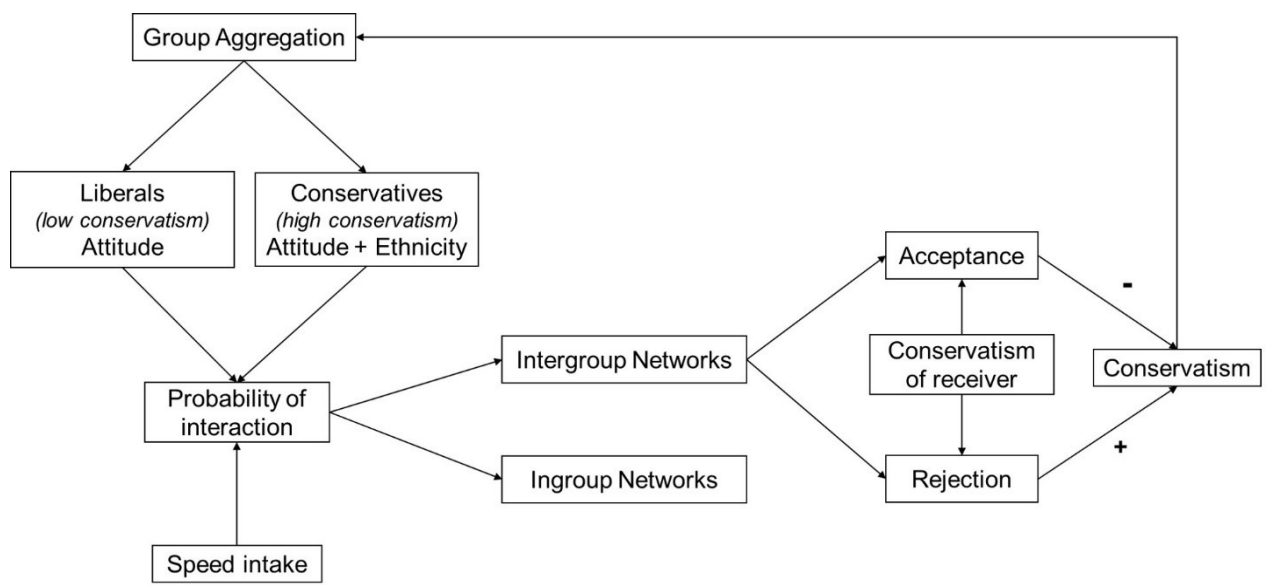

Figure 2. Prototype of MigrAgent model

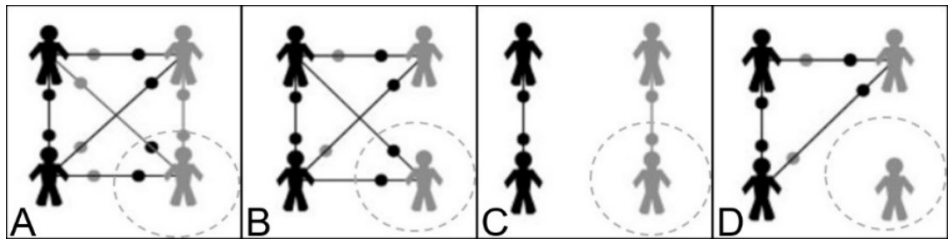

Figure 3. Stylized acculturation outcomes simulated in MigrAgent. Acculturation refers to the agent in dashed circle: A integration, B assimilation, C separation, D marginalization. Images are elaborated for grey-scale printing.

Experiments and Results

We used MigrAgent to explore the polarization dynamics and associated acculturation outcomes emerging from the interaction between a migrant and a local population in relation to network formation combining different levels of conservatism in both populations. We were additionally interested in how these outcomes varied depending on the speed of intake. To this end, we performed parameter sweeping, varying the parameters of the collective conservatism of the local population, the conservatism of the migrant population and the speed of intake. For the level of collective conservatism, we selected five conditions in both the migrant and the local population. The first condition was that of an equal distribution between liberals and conservatives (conservatism = 0). The other conditions were liberal societies (conservatism = -0.25), extremely liberal societies (conservatism = -0.75), conservative societies (conservatism = 0.25) and extremely conservative societies (conservatism = 0.75). For speed of intake, we selected condition 1 (minimum speed = 1% of the migrant population moves to the host country at each step) and 100 (maximal speed = 100% of the migrant population moves to the host country at each step). The number of agents was kept constant in all conditions: 500 local agents and 500 migrant agents. We computed a total of 50 combinations and counted 1,000 time steps for each simulation. Each simulation was repeated 20 times. Table 2 summarizes the parameter sweeps and experimental conditions. To avoid incremental effects of feedback in intergroup interactions, we put a limit to the change of conservatism of agent, equal to +1 for the upper level and -1 for the lower level. Agents change their individual level of conservatism as long it falls into the range. Outlier agents who fall out of the limit in the initial distribution at time step 0 do not change their individual level of conservatism, thus serving as noise to test the robustness of the model across the conditions. Results refer to the averaged scores out of the 20 repetitions.

| Parameters | Values | | | | |
|---|---|---|---|---|---|
| conservatism_local | -0.75 *Extremely liberal* | -0.25 *liberal* | 0 *equally distributed* | 0.25 *conservative* | 0.75 *Extremely conservative* |
| conservatism_migrant | -0.75 *extremely liberal* | -0.25 *liberal* | 0 *equally distributed* | 0.25 *conservative* | 0.75 *extremely conservative* |
| speed_intake | 1 *slow* | 100 *fast* | | | |
| number_local | 500 | | | | |
| number_migrant | 500 | | | | |
| Time steps | 1000 | | | | |
| Repetition runs | 20 | | | | |

Table 2. Parameter sweeps and experimental conditions

Describing our results, we first show a polarization effect at the macro-level of our observations. Then, we illustrate the acculturation processes along the simulation for substrata of society (liberal migrants, conservative migrants, liberal locals, conservative locals). Figure 4 serves as a baseline indicating the initial values of the simulation at time step 0 for each condition. Values reported in Figure 4 refer both to the local and migrant populations. Conservatism informs the mean level of conservatism in each population (i.e., how conservative or liberal the population is). The other two components of Figure 4 (rectangles) inform the fraction of liberals and migrants (i.e., how many agents of the category are in the population).

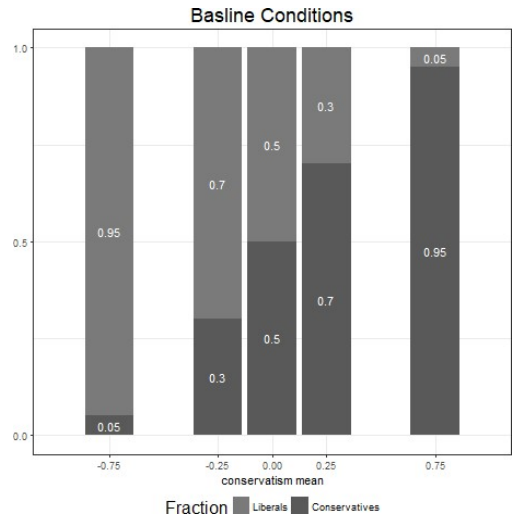

Figure 4. Baseline of conservatism mean, fraction of conservative agents and liberal agents at the start of the simulations (time step 0)

Figure 5 depicts the level of conservatism in each condition at the end of the simulation runs (final time step 1,000). The figures compare the migrant population (on the top) and local population (on the bottom), and for each observation, the slow intake speed and the fast intake speed are compared. Slow intake implies that approximately 250 migrants have entered the receiving society at time step 1,000, whereas fast intake means all 500 migrants have entered. Thus, the different intake speeds are a proxy for relative group sizes, with the migrant group in a minority condition in the slow intake condition. On the x-axis, the range of collective local conservatism at time step 0 is reported; on the y-axis, the range of collective migrant conservatism at time step 0 is reported. The cells in the heatmap report the average level of conservatism for that population. At first glance, for both of the populations, it is clear how polarization effects emerge with the tendency towards greater conservatism for conditions on the upper right-hand side of the diagonal and towards lower conservatism for conditions on the lower left-hand side of the diagonal. The heatmap provides an indication of how resilient one group is to the rejection of the other group. When more tolerant societies meet, they both shift to lower levels of conservatism (higher tolerance). Likewise, interactions between more conservative societies produce a shift towards greater conservatism for both. The local population is more sensitive to the degree of conservatism of the migrants than *vice versa*. The speed of intake of migrants has a stronger effect on the conservatism of the local population than on the migrants' degree of conservatism. For the migrant population, fast intake seems associated with a larger polarization effect. Figure 6 illustrates the acculturation outcomes that emerge at the global level for each population at time step 1,000. Values very close to 0 do not appear in the figures. The results highlight that integration and separation are the dominant scenarios. Conditions with convergence towards lower conservatism show integration as the dominant outcome, whereas for conditions with convergence towards greater conservatism, separation is the dominant outcome.

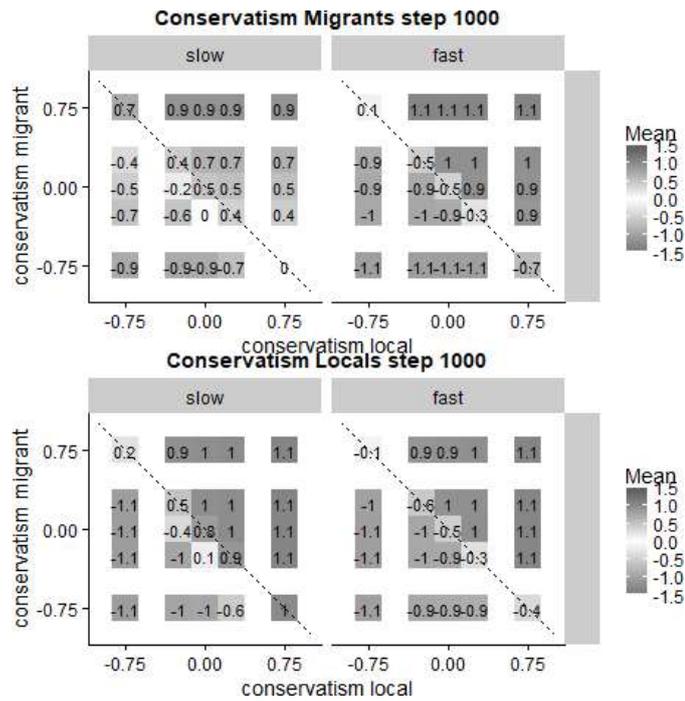

Figure 5. Conservatism mean at time step 1,000 for migrant and local population within slow and fast intake

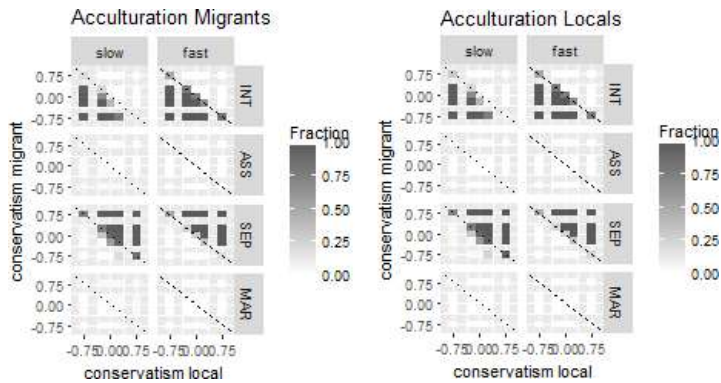

Figure 6. Acculturation outcomes at time step 1,000

Figure 7 reports the change in the fraction of liberals and conservatives in the migrant and local populations over the 1,000 time steps. Grey color refers to local population, black color refers to migrant population, while solid lines represent the fraction of liberal agents in each population and dotted lines conservative one.

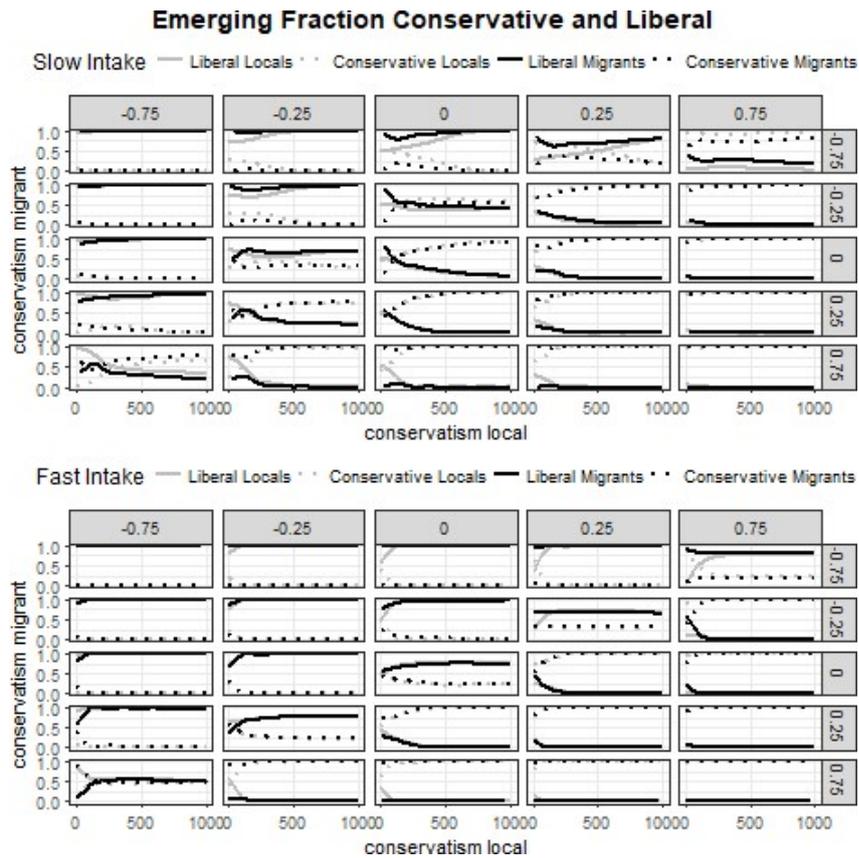

Figure 7. Emerging fraction liberal and conservative agents in local and migrant population over 1,000 time steps

In Figures 8,9,10 and 11, we explore the acculturation categories for each substratum of society over 1,000 time steps that can be compared with the change in the fraction of liberal or conservative agents as in Figure 7. The same graph is repeated for liberal migrants, conservative migrants, liberal locals and conservative locals. Each line represents the fraction of agents of the substratum that engages in that acculturation outcome.

As regards liberal migrants (Figure 8), assimilation is more evident in the early time steps of the simulation, but eventually it is overcome by integration. This outcome reflects how liberal migrants benefit from the availability of liberal locals they have more chance to interact with when in the condition of minority. The trend is more evident in the row for extremely conservative migrant population (conservatism = 0.75) in the slow intake condition. Comparing with Figure 7, we observe how integration becomes eventually the dominant outcome as long as at least a small part of liberals survives in both groups. When liberal agents turn conservative due to rejection by conservatives, surviving liberal migrants can still engage in assimilation for conditions of low conservatism in the local population (conservatism below 0). When the availability of liberal locals is not possible in the more conservative local population, separation emerges since liberal migrants can only find similar in their own ethnic group. In the fast intake condition, the change towards conservatism as effect of rejection is more drastic, with the extinction of liberal migrants when collective conservatism of either of the populations is above 0.

Conservative migrants (Figure 9) exhibit similar patterns but with opposite acculturation outcomes. In the early time steps of simulation runs, when conservative migrants are in the minority, marginalization arises for all combinations. The trend is more evident for slow intake conditions. This outcome happens because the conservative migrants, who are in the minority, refuse to interact with both the local population and liberal migrants. When more conservative agents enter the receiving society, separation is more prominent. In the fast intake condition, a trend toward separation emerges continuously and steadily already from the first time steps of the simulation. In the condition of extremely liberal societies (i.e., conservatism for both populations equal to -0.75), the conservative migrants disappear; they become liberals as effect of intergroup acceptance.

For liberal locals (Figure 10), an increase in separation appears for the first time steps, then overcome by integration. This result might seem counterintuitive, especially when extremely liberal migrant or extremely liberal local populations are involved. Similar to the trend of assimilation of liberal migrants, the reason is the unbalanced group sizes, which results in more opportunities for locals to interact with liberal members of the own group rather than migrants. As migrants enter the host society, integration steadily increases. When intake is fast, the convergence towards integration is more rapid, although at a lower rate than liberal migrants Interestingly, Figure 7 shows that the fractions of liberal migrants and liberal locals are almost identical. This similarity seems not reflected in the acculturation outcomes of the two populations comparing figure 8 and 10, and more evidently for the interaction of extremely liberal local population and extremely conservative migrant population. In this condition, liberal locals show integration at a much lower level than liberal migrants, both for slow intake and fast intake. The reason is likely to be again the relative group sizes. Since ingroup interactions occur by default, liberal locals have a higher probability to interact with other liberals of the own ethnic group than with liberal migrants. Additionally, although liberal locals would not aggregate with conservative locals, they have a high probability to relocate close to them due to the simple effect of larger group size. Therefore, diverse conditions favor segregation of liberal locals as simple effect of higher availability of co-ethnics.

Finally, the results for conservative locals (Figures 11) are not unexpected. They show more separation, and the trend follows the line of their distribution within the local population as shown in Figure 7. We can conclude that this outcome is a simple effect of unbalanced groups: the local population represents the majority, making it relatively easy for conservative locals to connect with others similar to themselves. As such, the lowest degree of separation for conservative locals is reached for conditions of fast intake and extremely liberal migrant societies, when they turn liberal because of the effect of intergroup acceptance

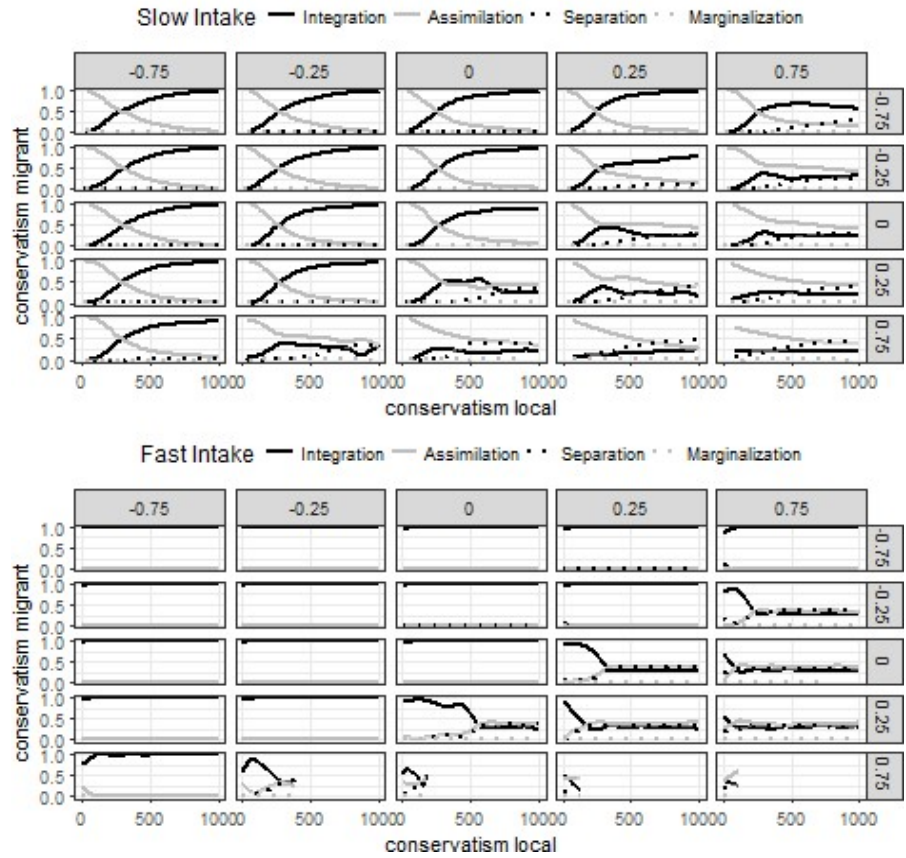

Figure 8. Acculturation outcomes for liberal migrants

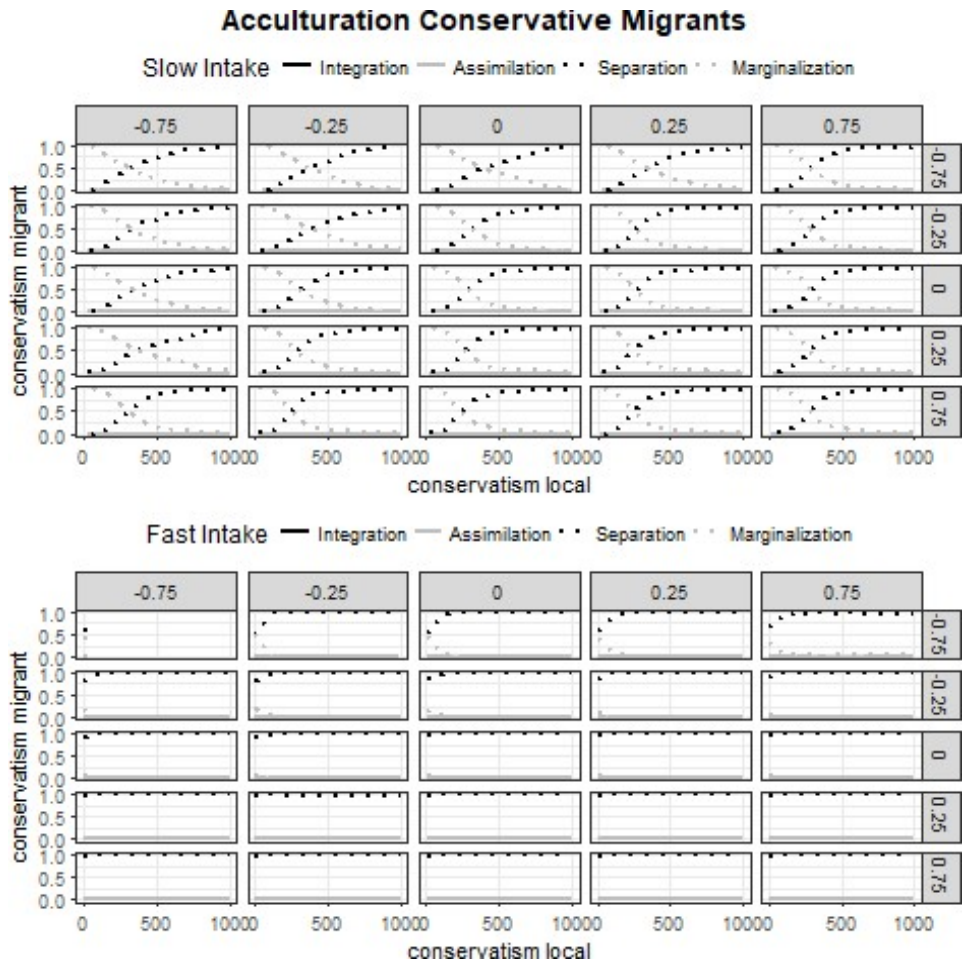

Figure 9. Acculturation outcomes for conservative migrants

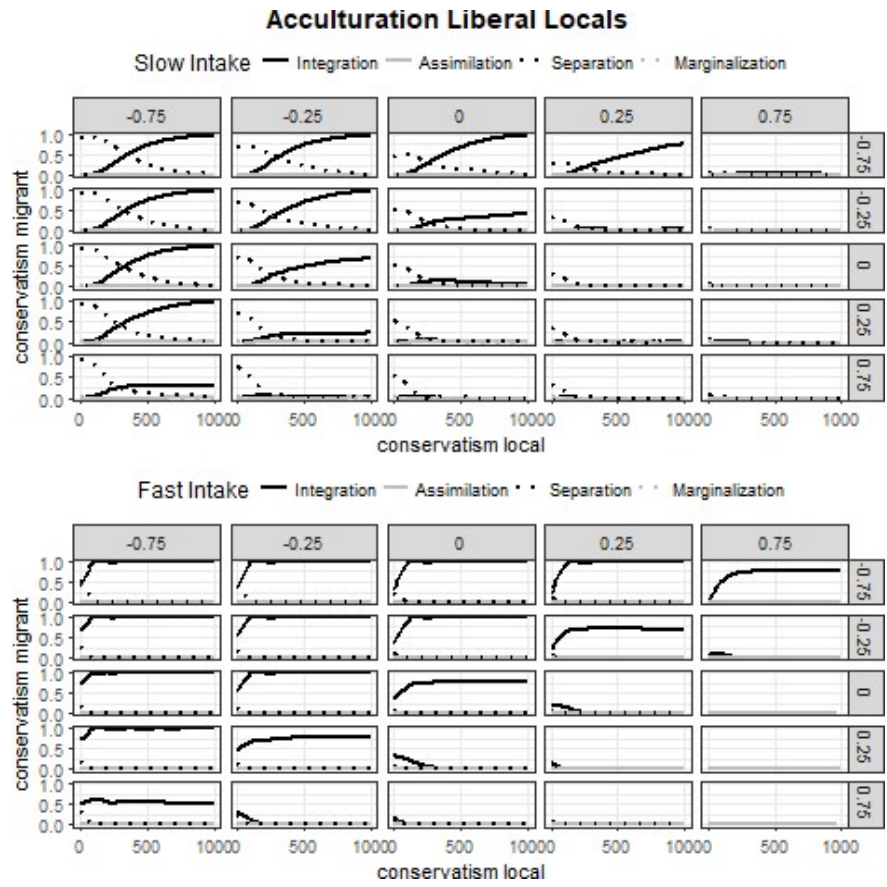

Figure 10. Acculturation outcomes for liberal locals

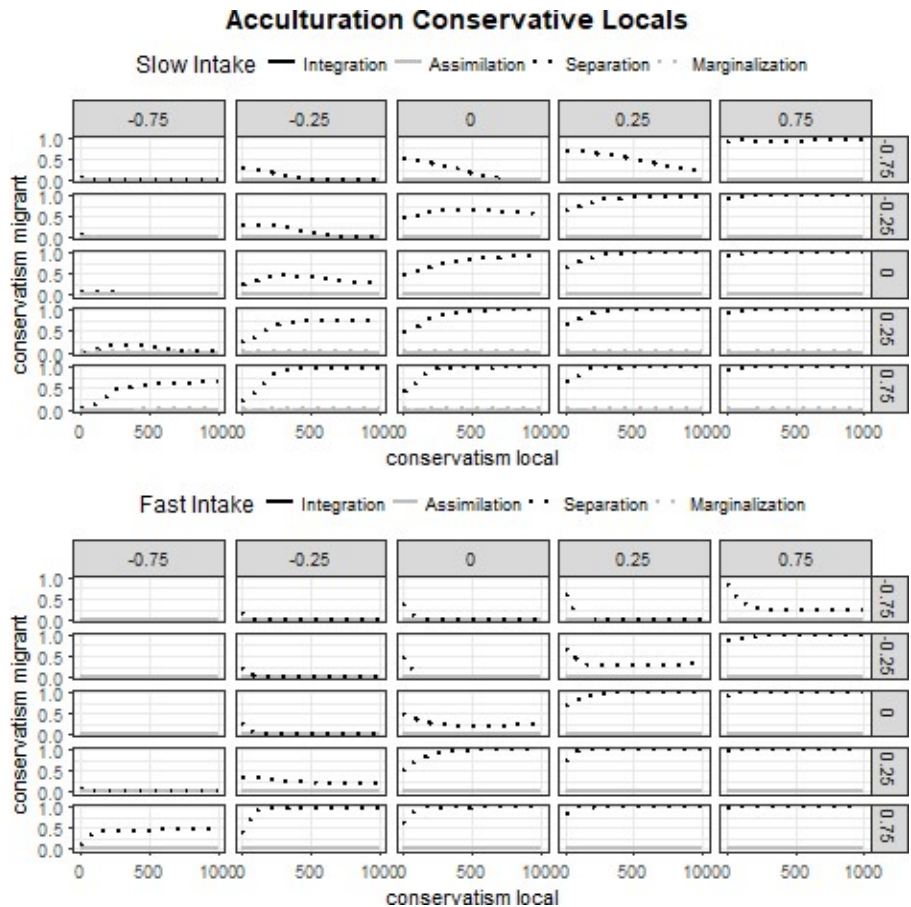

Figure 11. Acculturation outcomes for conservative locals

Our findings have implications regarding acculturation processes. First, the simulations demonstrate that different acculturation processes can coexist for different strata of society, and even within the same stratum. Slow intake influences acculturation processes through effects on relative group sizes. Acculturation strategies can emerge as a transitory stage, as seen with the liberal migrants' shift from assimilation to integration and the conservative migrants' shift from marginalization to separation. Counterintuitive scenarios, such as the separation of liberal locals in conditions where conservative agents represent the majority, are also possible. Furthermore, the simulations identified the effects of reciprocity due to acceptance and rejection between liberal and conservatives of the two populations. As for liberal migrants, reciprocity causes convergence to low levels of integration and assimilation in conditions of great conservatism among locals and migrants. Conversely, liberal locals who initially engage in integration shift to separation once they become conservative.

To check the effect sizes of the parameters involved in our simulations, we ran multiple linear regressions for each substratum of the population. Analyses were run in R. To ensure the readability of the text, a summary table of the multiple linear regressions is in the supplemental material. The dependent variables were the four acculturation outcomes. As for the predictors, we selected the emerged percentage of conservative locals, the percentage of conservative migrants, both over the 1,000 time steps, and the speed of intake as a factor variable. The regression models allowed us to compare the direction of causality of the predictors on the observed behavior and to calculate Cohen's $f^2$ for the effect sizes. The index is computed as $f^2 = \frac{sr^2}{1-R^2}$, where $sr^2$ is the squared semi-partial correlation and $R^2$ the multiple $R^2$ of the regression model. As a rule of thumb, we maintain that an $f^2 \geq 0.02$ denotes a small effect size, $f^2 \geq 0.15$ a medium effect size, and $f^2 \geq 0.35$ a large effect size (Kabacoff, 2011). The results are noteworthy for the consistent acculturation outcomes in each subgroup, as observed in Figures 8, 9, 10 and 11. As regards the integration of liberal migrants, the regression models confirm a negative effect of the percentage of conservative locals and conservative migrants, which are the groups that tend to isolate liberal migrants. The speed of intake is positively related to the integration of liberal migrants and negatively related to their assimilation. The Cohen's $f^2$ for intake speed is $f^2 = 0.28$ for integration of liberal migrants and $f^2 = 0.32$ for assimilation of liberal migrants, close to a large effect size. As for conservative migrants, the percentage of conservative locals is negatively related to separation and positively related to marginalization. In contrast, the percentage of conservative migrants is positively linked to separation and negatively linked to marginalization.

The effect of conservative locals on separation of conservative migrants seems counterintuitive, and it might depend on the role of the other types of agents and reciprocity effects. On the contrary, the other results fit the theoretical assumptions of the model. Increased number of conservative locals favors their chance to cluster, with the exclusion of conservative migrants and the increase in their marginalization. At the same time, the large number of conservative migrants increases the chance of interaction with other similar migrants, with the result to increase separation and buffer against marginalization. As for speed of intake, larger group size is positively associated with the separation of conservative migrants and negatively associated with their marginalization, because of the influence on the probability of interaction. For both cases, Cohen's $f^2$ indicates a large effect size ($f^2 = 0.40$). As for liberal locals, an interesting result is that an increased percentage of conservative locals is negatively related to integration at a high rate, due to the rejection of liberals by conservatives because of the different attitude. The speed of intake exhibits large effect sizes for integration ($f^2 = 1.72$) and marginalization ($f^2 = 0.43$), although $R^2$ for integration outcome is suspiciously high and more likely due to a simple effect. Finally, for conservative locals, the observations can be interpreted as a simple effect of their distribution, and indeed, the results of regression models are spurious.

## Comparison with Empirical Data

We qualitatively compared the assumptions and observations of our model with data from the second wave of the *Causes and Consequences of Socio-Cultural Integration Processes among New Immigrants in Europe* (SCIP) survey in Germany (Diel et al., 2015). The survey is a panel study interested in the antecedents of the adaptation of first-generation migrants in four European countries: Germany, the Netherlands, Ireland, and the UK. We focused on the second wave (2012/2013) after one year of residence in the country. Although one year of adaptation might not be a long enough time period when considering acculturation, the items of the survey seemed to be appropriate proxies for the parameters of our model. We selected the German sample because of consistency in the number of participants as compared to other countries' samples; moreover, the data collected best fit our model parameters. After approximating normality assumptions to run multiple linear regression models, we were left with a final sample of 1,224 migrants of Polish and Turkish origins. Analysis can be reproduced via the R-code provided.

Table 3 illustrates the items we selected as proxies for the parameters in our model and their ranges. As for the typology of networks as the dependent variable, we selected the frequency of interactions with locals and co-ethnics. As for the predictors in our model, we selected the experience of rejection as a proxy for rejection in the host country, the experience of hospitality as a proxy for acceptance and the importance of the country of birth as a proxy for the conservatism of individuals.

| Proxy for | Item | Range Scale |
| --- | --- | --- |
| Conservatism | IPIDCB: How important is the following to your sense of who you are: the country where you were born? | 1: very important 4: not important at all |
| Rejection | DSCRFREQ: […] How often do you think CO people are discriminated against in RC? | 1: very often 5: never |
| Acceptance | HOSP_RC: In general, RC is a hospitable/welcoming country for [CO people/pl]? | 1: strongly agree 5: strongly disagree |
| Interactions with locals | PPRC: How often do you spend time with [RC people/pl]? | 1: every day 6: never |
| Interactions with co-ethnics | PPCO: How often do you spend time with [CO people/pl]? | 1: every day 6: never |
| RC: receiving country; CO: country of origin | | |

Table 3. Items from SCIP survey used as proxy in our analyses

We ran multiple linear regressions to detect the direction of causality between the predictors and types of networks and then checked if the results fit the observed processes of our model. For the interpretation of the results, one should refer to the scale range of the items in Table 3. As for interactions with locals (Table 4), the negative coefficient *b* for conservatism implies that the more important participants considered their country of birth (1 = very important) in defining themselves (proxy for conservatism), the fewer interactions they had with locals of other ethnicity (6 = never). Conversely, the less important people considered their country of birth in defining themselves (4 = not important), the more likely they were to interact with locals (1 = every day). As for rejection, the same interpretation of the negative coefficient *b* applies. People who reported high levels of rejection (1 = very often) were less likely to interact with locals (6 = never), whereas people who reported lower levels of rejection (5 = never) were more likely to interact with locals (1 = every day). The same held true for the positive coefficient *b* for acceptance: Migrants who felt accepted in the host country (1 = strongly agree) were more likely to connect with locals (1 = every day), whereas people who did not feel accepted (5 = strongly disagree) were less likely to connect with them (6 = never).

As for interactions with co-ethnics (Table 5), the positive direction of coefficient *b* for conservatism fits our model, meaning that migrants considering their country of birth as very important to their self-definition (1 = very important) were more likely to interact with co-ethnics (1 = every day), and those who did not value their country of birth as much (4 = not important at all) were less likely to interact with co-ethnics (6 = never). We can expect rejection by the local population to be positively related to interactions with co-ethnics, as happens in our model with segregation patterns due to effects of intergroup rejection. Although coefficient *b* is positive and effect size strong ($f^2 = 0.57$), its magnitude is close to 0 and p-value not significant, so that our hypothesis cannot be confirmed. As regards the effect of acceptance by locals on interactions with co-ethnics, we cannot formulate firm hypothesis considering ingroup interactions happen by default in our model, and unless attitude of migrants and spatial sorting are taken into consideration. As a matter of fact, coefficient *b* for experience of acceptance by locals denotes a null correlation, and a not significant p-value.

In sum, the regression analyses confirmed the plausibility of our model's results, in particular for interactions with locals with significant results. Yet, the magnitude of coefficient *b*, as compared to the simple regression and partial $R^2$, was not very large. Namely for interactions with co-ethnics, the results did not strongly support our model. Additionally, the effect sizes (Cohen's $f^2$) were strong for the effects of rejection and acceptance on interactions with locals and co-ethnics. Nevertheless, predicted interactions with co-ethnics were not associated with significant p-values for all predictors, except for the proxy for conservatism.

| Interactions with locals | | | | | | |
|---|---|---|---|---|---|---|
| Predictor | b | partial R² | r | sr² | Cohen f² | p-value |
| conservatism | -0.275 | 0.020 | -0.15 | 0.08 | 0.09 | 3.01e-06*** |
| experience rejection | -0.169 | 0.012 | -0.14 | 0.63 | 0.66 | 0.0017** |
| experience acceptance | 0.126 | 0.008 | 0.12 | 1.27 | 1.33 | 0.0204* |
| (Intercept) | 2.812 | | | | | < 2e-16*** |
| R² | 0.046 | | | | | 3.089e-10*** |
| Adj R² | 0.043 | | | | | |
| Signif. codes: 0 '***' 0.001 '**' 0.01 '*' 0.05 '.' 0.1 ' ' 1 | | | | | | |

Table 4. Summary multiple linear regression model with SCIP data for interactions with locals

| Interactions with co-ethnics | | | | | | |
|---|---|---|---|---|---|---|
| Predictor | b | partial R² | r | sr² | Cohen f² | p-value |
| conservatism | 0.179 | 0.013 | 0.11 | 0.00 | 0.01 | 0.000368*** |
| experience rejection | 0.051 | 0.001 | 0.04 | 0.56 | 0.57 | 0.268089 |
| experience acceptance | 0.018 | -0.002 | 0.00 | 0.45 | 0.46 | 0.701909 |
| (Intercept) | 1.640 | | | | | 1.83e-12*** |
| R² | 0.014 | | | | | 0.002499** |
| Adj R² | 0.011 | | | | | |
| Signif. codes: 0 '***' 0.001 '**' 0.01 '*' 0.05 '.' 0.1 ' ' 1 | | | | | | |

Table 5. Summary multiple linear regression model with SCIP data for interactions with co-ethnics

## Limitations and Conclusions

In this preliminary work, we have presented our MigrAgent model and demonstrated how different acculturation processes can emerge and coexist for different groups in society in a stylized scenario. We have also explored these processes' association with polarization dynamics. We think our work shows that agent-based modelling can be a valuable contribution to the study of acculturation, but we recognize much must be done to increase the validity of our model. A main challenge in the work presented herein was the comparison with real data. We recognize that although the selected empirical data reasonably support our model, they are not very strong for multiple reasons. First of all, these were secondary data we adopted as proxies for our parameters, and so the meanings of the items might not fit those of the model. This could explain the lack of consistency for interactions with locals. Even if we could access data fitting the meanings of the parameters of our model, we would need historical data to validate the processes in MigrAgent. Additionally, the sample of SCIP do not include locals, who are a class of actors included in our model. A survey specifically designed to feed our simulation might be envisioned in the future, testing the hypothesis generated by the processes observed in MigrAgent and addressing specifically the interaction between locals and migrants over time. On the theoretical level, one issue regards the definition of acculturation. Migrant integration is a complex phenomenon that can be studied from a psychological, structural or intergenerational perspective (Rumbaut, 2015). At this stage, our model allows us to differentiate between an individual and a collective level, which is the main purpose of agent-based modelling (Squazzoni, Jager, & Edmonds, 2013). Although this feature can be an asset in the study of acculturation processes, it is still necessary to consider what specific areas of the literature might benefit from the model and how its contribution can be best oriented. Here, we have focused on networks in terms of the participation of individuals from different cultures (Berry, 2005). This approach fits our research question, but many other dimensions should be included, depending on the definition of acculturation. A critical one is the dimension of power. The asymmetrical power relations between locals and migrants have been widely recognized (Verkuyten, Yogeeswaran, & Adelman, 2018), which justifies the focus on exploring the dynamics of opinion polarization in receiving countries. However, adding a dimension of power would imply the inclusion of additional parameters and different mechanisms of interaction between the two populations. For the sake of the model's simplicity at this stage, we avoided such considerations in this study. Finally, the results of our simulations illustrate strong polarization between integration and separation, which might have obscured more nuanced marginalization and assimilation conditions. We theorize these results were due to the choice to keep the desired fraction of similar others at 0.50, which means that agents would relocate as soon as those considered as similar are in the minority condition. This scenario increases chances of spatial segregation of agents and consequently influences formation of agents' network in their proximity. Nevertheless, our parameter sweeping already included too many conditions to compare, and we opted for this choice. Sweeping the fraction of desired others would be a feasible area of investigation in the future, so to better understand how spatial sorting influences the network formation. Further studying could also include thorough tests on the robustness of the model and boundary effects on the distribution of conservatism in the groups.

Even recognizing certain limits due to the preliminary status of our work, we think MigrAgent can provide a useful basis for the study of acculturation dynamics and the conditions under which mutual acceptance can emerge or fail, resulting in different scenarios for societal network structures. The main contribution of the model is that it compares different types of societies and strata for multiple groups, in this case along the dimension of conservatism/liberalism. In the future, societies or their strata might differ along other dimensions, such as the internal distribution of wealth within groups (Bergreen & Nilsson, 2003) or the similarity of cultural models and practices (Hofstede, 2001). An asset of MigrAgent is its ability to translate these social dynamics into scenarios linking migration flows and post-migration adaptation. We aim in the future to parameterize MigrAgent using available data to further improve its empirical grounding and conduct studies on specific cases of different migration scenarios.

Author Information

Rocco Paolillo, M.A., is a EU COFUND BIGSSS-departs PhD fellow at the Bremen International Graduate School of Social Sciences, University of Bremen & Jacobs University Bremen (Germany). His work for the project here presented is related to a research stay at the University College Groningen (the Netherlands) funded by the Foundation "Franco e Marilisa Caligara" in Torino (Italy) in 2015 and with the support of University College Groningen.

email: rpaolillo@bigsss-bremen.de

Wander Jager, Ph.D., is an associate professor of Social Complexity at University College Groningen and managing director of the Groningen Center for Social Complexity Studies, both at the University of Groningen (the Netherlands).

email: w.jager@rug.nl

Additional Information

NetLogo model and R-code for data analysis can be found at:

Paolillo, Rocco, Jager, Wander (2018, November 28). "MigrAgent" (Version 1.2.0). *CoMSES Computational Model Library*. Retrieved from: https://www.comses.net/codebases/a6fc3cc6-bd5b-4cd6-8300-75978ea1e362/releases/1.2.0/.

Empirical data from SCIP project is available at GESIS Data Archive (study nr. ZA5956):

Diehl, C., Gijsberts, M., Güveli, A., Koenig, M., Kristen, C., Lubbers, M., McGinnity, F., Mühlau, P., Platt, L., & Van Tubergen, F. (2016). Causes and consequences of socio-cultural integration processes among new immigrants in Europe (SCIP). GESIS Data Archive, Cologne. ZA5956 Data file Version 1.0.0, doi:10.4232/1.12341. Retrieved from https://dbk.gesis.org/dbksearch/sdesc2.asp?no=5956&db=e&doi=10.4232/1.12341

# Supplemental Material

Multiple Linear Regression Models Simulations

R²: Multiple R²; Adj R²: Adjusted R²; b: regression coefficient; r: correlation coefficient; sr²: squared semi-partial correlation; Cohen f²: effect size

| Integration Liberal Migrants | | | | | | |
|---|---|---|---|---|---|---|
| Predictor | b | partial R² | r | sr² | Cohen f² | p-value |
| % conservative locals | -0.363 | 0.016 | -0.67 | 0.00 | 0.00 | <2e-16 *** |
| % conservative migrants | -0.151 | 0.003 | -0.66 | 0.00 | 0.00 | <2e-16 *** |
| Speed intake | 0.197 | 0.130 | 0.42 | 0.12 | 0.28 | <2e-16 *** |
| (Intercept) | 0.809 | | | | | <2e-16 *** |
| R² | 0.570 | | | | | |
| Adj R² | 0.570 | | | | | |

Signif. codes: 0 '***' 0.001 '**' 0.01 '*' 0.05 '.' 0.1 ' ' 1

| Assimilation Liberal Migrants | | | | | | |
|---|---|---|---|---|---|---|
| Predictor | b | partial R² | r | sr² | Cohen f² | p-value |
| % conservative locals | 0.197 | 0.005 | 0.47 | 0.00 | 0.00 | < 2e-16 *** |
| % conservative migrants | 0.065 | 0.001 | 0.45 | 0.00 | 0.00 | 3.72e-07 *** |
| Speed intake | -0.252 | 0.212 | -0.51 | 0.20 | 0.32 | < 2e-16 *** |
| (Intercept) | 0.254 | | | | | < 2e-16 *** |
| R² | 0.388 | | | | | |
| Adj R² | 0.388 | | | | | |

Signif. codes: 0 '***' 0.001 '**' 0.01 '*' 0.05 '.' 0.1 ' ' 1

Multiple Linear Regression Models Simulations

R²: Multiple R²; Adj R²: Adjusted R²; b: regression coefficient; r: correlation coefficient; sr²: squared semi-partial correlation; Cohen f²: effect size

| Separation Liberal Migrants | | | | | | |
|---|---|---|---|---|---|---|
| Predictor | b | partial R² | r | sr² | Cohen f² | p-value |
| % conservative locals | 0.158 | 0.011 | 0.58 | 0.00 | 0.00 | <2e-16 *** |
| % conservative migrants | 0.088 | 0.003 | 0.58 | 0.00 | 0.00 | <2e-16 *** |
| Speed intake | 0.061 | 0.048 | 0.04 | 0.04 | 0.07 | <2e-16 *** |
| (Intercept) | -0.066 | | | | | <2e-16 *** |
| R² | 0.374 | | | | | |
| Adj R² | 0.374 | | | | | |

Signif. codes: 0 '***' 0.001 '**' 0.01 '*' 0.05 '.' 0.1 ' ' 1

| Marginalization Liberal Migrants | | | | | | |
|---|---|---|---|---|---|---|
| Predictor | b | partial R² | r | sr² | Cohen f² | p-value |
| % conservative locals | 0.008 | 0.001 | 0.13 | 0.00 | 0.00 | 1.28e-09 *** |
| % conservative migrants | -0.002 | 0.000 | 0.12 | 0.00 | 0.00 | 0.0883 . |
| Speed intake | -0.005 | 0.012 | -0.14 | 0.01 | 0.01 | < 2e-16 *** |
| (Intercept) | 0.004 | | | | | < 2e-16 *** |
| R² | 0.029 | | | | | |
| Adj R² | 0.029 | | | | | |

Signif. codes: 0 '***' 0.001 '**' 0.01 '*' 0.05 '.' 0.1 ' ' 1

Multiple Linear Regression Models Simulations

R²: Multiple R²; Adj R²: Adjusted R²; b: regression coefficient; r: correlation coefficient; sr²: squared semi-partial correlation; Cohen f²: effect size

| Separation Conservative Migrants | | | | | | |
|---|---|---|---|---|---|---|
| Predictor | b | partial R² | r | sr² | Cohen f² | p-value |
| % conservative locals | -0.117 | 0.002 | 0.01 | 0.00 | 0.00 | <2e-16 *** |
| % conservative migrants | 0.189 | 0.005 | 0.03 | 0.00 | 0.00 | <2e-16 *** |
| Speed intake | 0.312 | 0.294 | 0.54 | 0.28 | 0.40 | <2e-16 *** |
| (Intercept) | 0.634 | | | | | <2e-16 *** |
| R² | 0.303 | | | | | |
| Adj R² | 0.303 | | | | | |

Signif. codes: 0 '***' 0.001 '**' 0.01 '*' 0.05 '.' 0.1 ' ' 1

| Marginalization Conservative Migrants | | | | | | |
|---|---|---|---|---|---|---|
| Predictor | b | partial R² | r | sr² | Cohen f² | p-value |
| % conservative locals | 0.117 | 0.002 | -0.01 | 0.00 | 0.00 | <2e-16 |
| % conservative migrants | -0.189 | 0.005 | -0.03 | 0.00 | 0.00 | <2e-16 |
| Speed intake | -0.312 | 0.294 | -0.54 | 0.28 | 0.40 | <2e-16 |
| (Intercept) | 0.366 | | | | | <2e-16 |
| R² | 0.303 | | | | | |
| Adj R² | 0.303 | | | | | |

Signif. codes: 0 '***' 0.001 '**' 0.01 '*' 0.05 '.' 0.1 ' ' 1

Multiple Linear Regression Models Simulations

$R^2$: Multiple $R^2$; Adj $R^2$: Adjusted $R^2$; b: regression coefficient; r: correlation coefficient; $sr^2$: squared semi-partial correlation; Cohen $f^2$: effect size

| Integration Liberal Locals | | | | | | |
|---|---|---|---|---|---|---|
| Predictor | b | partial $R^2$ | r | $sr^2$ | Cohen $f^2$ | p-value |
| % conservative locals | -0.803 | 0.19 | -0.92 | 0.01 | 0.06 | < 2e-16 *** |
| % conservative migrants | -0.059 | 0.11 | -0.90 | 0.00 | 0.00 | 3.03e-15 *** |
| Speed intake | 0.153 | 0.22 | 0.35 | 0.21 | 1.72 | < 2e-16 *** |
| (Intercept) | 0.774 | | | | | < 2e-16 *** |
| $R^2$ | 0.881 | | | | | |
| Adj $R^2$ | 0.881 | | | | | |

Signif. codes: 0 '***' 0.001 '**' 0.01 '*' 0.05 '.' 0.1 ' ' 1

| Assimilation Liberal Locals | | | | | | |
|---|---|---|---|---|---|---|
| Predictor | b | partial $R^2$ | r | $sr^2$ | Cohen $f^2$ | p-value |
| % conservative locals | -0.001 | 0.00 | -0.29 | 0.00 | 0.00 | 1.14e-12 *** |
| % conservative migrants | 0.000 | 0.00 | -0.28 | 0.00 | 0.00 | 2.28e-06 *** |
| Speed intake | 0.001 | 0.05 | 0.25 | 0.04 | 0.05 | < 2e-16 *** |
| (Intercept) | 0.002 | | | | | < 2e-16 *** |
| $R^2$ | 0.124 | | | | | |
| Adj $R^2$ | 0.124 | | | | | |

Signif. codes: 0 '***' 0.001 '**' 0.01 '*' 0.05 '.' 0.1 ' ' 1

Multiple Linear Regression Models Simulations

R²: Multiple R²; Adj R²: Adjusted R²; b: regression coefficient; r: correlation coefficient; sr²: squared semi-partial correlation; Cohen f²: effect size

| Separation Liberal Locals | | | | | | |
|---|---|---|---|---|---|---|
| Predictor | b | partial R² | r | sr² | Cohen f² | p-value |
| % conservative locals | -0.197 | 0.01 | -0.27 | 0.00 | 0.00 | < 2e-16 *** |
| % conservative migrants | 0.060 | 0.00 | -0.27 | 0.00 | 0.00 | 8.74e-16 *** |
| Speed intake | -0.153 | 0.21 | -0.39 | 0.20 | 0.28 | < 2e-16 *** |
| (Intercept) | 0.223 | | | | | < 2e-16 *** |
| R² | 0.271 | | | | | |
| Adj R² | 0.271 | | | | | |

Signif. codes: 0 '***' 0.001 '**' 0.01 '*' 0.05 '.' 0.1 ' ' 1

| Marginalization Liberal Locals | | | | | | |
|---|---|---|---|---|---|---|
| Predictor | b | partial R² | r | sr² | Cohen f² | p-value |
| % conservative locals | 0.000 | 0.00 | -0.16 | 0.00 | 0.00 | <2e-16 *** |
| % conservative migrants | -0.001 | 0.33 | -0.18 | 0.00 | 0.00 | <2e-16 *** |
| Speed intake | -0.001 | 0.30 | -0.51 | 0.28 | 0.43 | <2e-16 *** |
| (Intercept) | 0.001 | | | | | <2e-16 *** |
| R² | 0.331 | | | | | |
| Adj R² | 0.331 | | | | | |

Signif. codes: 0 '***' 0.001 '**' 0.01 '*' 0.05 '.' 0.1 ' ' 1

Multiple Linear Regression Models Simulations

R²: Multiple R²; Adj R²: Adjusted R²; b: regression coefficient; r: correlation coefficient; sr²: squared semi-partial correlation; Cohen f²: effect size

| Separation Conservative Locals | | | | | | |
|---|---|---|---|---|---|---|
| Predictor | b | partial R² | r | sr² | Cohen f² | p-value |
| % conservative locals | 0.995 | 1.000 | 1.00 | 0.04 | NA | <2e-16 *** |
| % conservative migrants | -0.002 | 0.994 | 0.98 | 0.00 | NA | <2e-16 *** |
| Speed intake | -0.002 | 0.257 | -0.18 | 0.24 | NA | <2e-16 *** |
| (Intercept) | 0.000 | | | | | <2e-16 *** |
| R² | 1.000 | | | | | |
| Adj R² | 1.000 | | | | | |

Signif. codes: 0 '***' 0.001 '**' 0.01 '*' 0.05 '.' 0.1 ' ' 1

| Marginalization Conservative Locals | | | | | | |
|---|---|---|---|---|---|---|
| Predictor | b | partial R² | r | sr² | Cohen f² | p-value |
| % conservative locals | 0.005 | 0.079 | 0.87 | 0.00 | 0.02 | <2e-16 *** |
| % conservative migrants | 0.002 | 0.817 | 0.87 | 0.00 | 0.00 | <2e-16 *** |
| Speed intake | 0.002 | 0.257 | 0.09 | 0.24 | 1.38 | <2e-16 *** |
| (Intercept) | 0.000 | | | | | <2e-16 *** |
| R² | 0.822 | | | | | |
| Adj R² | 0.822 | | | | | |

Signif. codes: 0 '***' 0.001 '**' 0.01 '*' 0.05 '.' 0.1 ' ' 1